\documentclass{appolb}
\usepackage{epsfig}
\usepackage{axodraw}

\begin{document}
\title{Multiple Interactions, Saturation, and Final States in $pp$ 
Collisions and DIS
\thanks{Presented at the Epiphany Conference, Cracow, January 2009}}
\author{G\"osta Gustafson\\
\address{Dept. of Theoretical Physics, Lund Univ., Sweden\\
and II Inst. f\"{u}r Theor. Physik, Hamburg Univ., Germany\\
e-mail: gosta.gustafson@thep.lu.se}
}
\maketitle
\begin{abstract}
\emph{Dedicated to the memory of Jan Kwieci\'{n}ski}

\vspace{2mm}
In high energy collisions saturation and multiple collisions are most easily
accounted for in transverse coordinate space, while analyses in momentum space 
have been more suitable for calculating properties of exclusive final states. 
In this talk I describe an extension of Mueller's dipole cascade model, which
attempts to combine the good features of both these descriptions. Besides 
saturation it also includes effects of correlations and fluctuations, which 
have been difficult to account for in previous approaches. The model
reproduces successfully total, elastic, and diffractive cross sections
in $pp$ collisions and DIS, and a description of final states will be ready
soon.
\end{abstract}

\section{Introduction}

In this talk I want to discuss some results related to saturation and multiple 
collisions obtained in 
collaboration with Emil Avsar, Christoffer Flensburg, and Leif L\"onnblad
\cite{newmodel, elastic}.
In high energy $pp$ collisions the minijet cross section is much larger than 
the total cross section. This implies that each event contains on average 
more than one minijet pair, and multiple subcollisions are an essential 
feature of high energy hadronic reactions.
A formulation in transverse coordinate space and the eikonal approximation
are particularly suited for a treatment of these features,
and has been successfully applied to $\gamma^* p$ collisions, both for total 
and diffractive scattering cross sections \cite{GBW, BGBK}.

The application of the eikonal formalism is, however, mainly applicable for 
inclusive observables, and a description of exclusive final states is more 
easy in momentum space. At present the most successful model for high energy 
$pp$ collisions is the \textsc{Pythia} model by Sj\"ostrand and coworkers
\cite{Pythia},
which has been extensively tuned to Tevatron data by R. Field \cite{Field}.

There are, however, a set of problems connected to these tunes. How does the 
partonic final state hadronize? A fit to data needs strong colour 
reconnections, for which we have a poor theoretical understanding. Also it is 
difficult to properly include effects of correlations and fluctuations in 
momentum space cascades.

In this talk I want to discuss a new approach based on Mueller's dipole 
cascade model,
which is an attempt to combine the good features of formulations in 
transverse coordinate space and in momentum space.

\section{Minimum bias and underlying events}

\begin{figure}
\begin{center}
\includegraphics[scale=0.4]{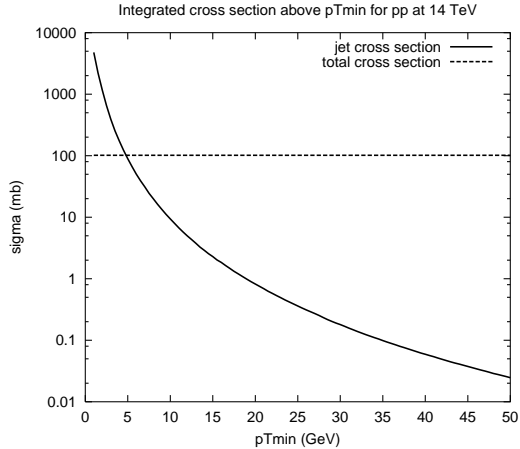}
\end{center}
\caption{Integrated jet cross section above a cut $p_{\perp,\mathrm{min}}$ 
at 14 TeV, from ref. \cite{jetcrossection}.}
\label{fig:ptint}
\end{figure}

In collinear factorization the inclusive parton scattering cross section 
diverges like $d \hat{\sigma} / d p_\perp^2 \sim 1/p_\perp^4$ for small
$p_\perp$. 
An estimate of the integrated jet cross section above a cut 
$p_{\perp,\mathrm{min}}$
is shown in fig.~\ref{fig:ptint} for $pp$ scattering at 14 TeV. We here see 
that for $p_{\perp,\mathrm{min}}<5$ GeV the 
integrated jet cross section exceeds the total cross section, which
implies that an average collision must have several hard subcollisions.

Multiple collisions have also been directly observed in collider experiments.
Thus four-jet events with pairwise back-to-back jets have been seen
by AFS at the ISR, by UA2 at the CERN S$p\bar{p}$S collider, and by CDF at 
the Tevatron. At the Tevatron the most clear signal is observed in the
analysis of 3 jets + a prompt photon \cite{CDFjetgamma}.

It was early suggested that the increase in the $pp$ cross section
is driven by minijet production, and that minimum bias events are dominated by
(semi)hard parton subcollisions \cite{Cline}. This is also an essential 
assumption in the \textsc{Pythia} model. This model is also based on
collinear factorization, which implies that a cutoff is needed for
the singularity at small $p_\perp$. This is motivated by the fact that 
hadrons are colour neutral, and therefore the Coulomb potential must be 
screened for large impact parameters or small $p_\perp$. Fits to collider 
data give a cutoff at around 2 GeV, slowly increasing at higher energies.
(A similar cutoff is also obtained naturally in the $k_\perp$-factorization 
formalism \cite{Miu}. In the dipole cascade model the transverse
momentum $k_\perp$ of a colliding
gluon is related to the dipole size in transverse coordinate space, 
and therefore also to the screening length.)

For the more recent version of \textsc{Pythia} this cutoff gives typically
2 - 3 interactions per event at the Tevatron, and 4 - 5 at the LHC.
An important feature is also that the subcollisions are correlated.
Central collisions have many interactions, while peripheral collisions 
have few. In the experimental analyses this is described in terms of an 
effective cross section $\sigma_{\mathrm{eff}}$. The cross section 
$\sigma_{\mathrm{DPS}}$ for the simultaneous
hard reactions $A$ and $B$ (with $A \ne B$) is written as
$\sigma_{\mathrm{DPS}} = \sigma_A \sigma_B / \sigma_{\mathrm{eff}}$.
If the interactions $A$ and $B$ were uncorrelated we would have 
$\sigma_{\mathrm{eff}}$ equal to the inelastic nondiffractive cross section
($\sim$ 50 mb at the Tevatron).
Instead the experimental results give the much smaller result
$\sigma_{\mathrm{eff}} \approx
15$ mb, which thus corresponds to a very strong enhancement.

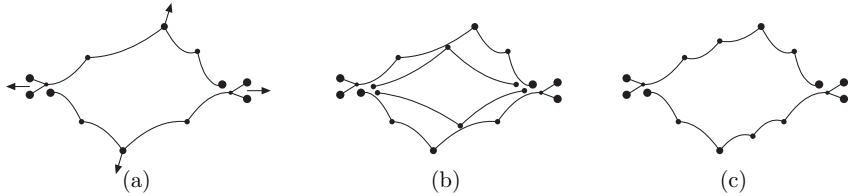
\begin{figure}
\scalebox{0.78}{\mbox{
\begin{picture}(301,115)(34,20)
\Curve{(73,85)(86,87)(110,100)}
\Curve{(90,40)(110,53)(121,54)}
\Curve{(223,85)(236,87)(260,100)}
\Curve{(240,40)(260,53)(271,54)}
\Curve{(211,71)(224,75)(247,90)}
\Curve{(247,90)(268,75)(280,72)}
\Curve{(213,68)(227,65)(253,52)}
\Curve{(253,52)(273,66)(284,69)}

\LongArrow(110,100)(113,110)
\LongArrow(90,40)(87,30)
\LongArrow(150,69)(160,69)
\LongArrow(45,71)(35,71)

\Curve{(53,72)(63,75)(73,85)}
\Curve{(110,100)(115,92)(126,88)}
\Curve{(126,88)(127,84)(133,73)(138,72)}
\Vertex(73,85){1.3}
\Vertex(110,100){1.7}
\Vertex(126,88){1.3}
\Vertex(138,72){2}
\Line(53,72)(45,75)
\Line(53,72)(45,67)
\Vertex(53,72){1}
\Vertex(45,75){2}
\Vertex(45,67){2}

\Curve{(55,68)(63,65)(70,54)}
\Curve{(70,54)(82,49)(90,40)}
\Curve{(121,54)(130,64)(142,68)}
\Vertex(70,54){1.3}
\Vertex(90,40){1.7}
\Vertex(121,54){1.3}
\Line(142,68)(150,73)
\Line(142,68)(150,65)
\Vertex(142,68){1}
\Vertex(150,73){2}
\Vertex(150,65){2}
\Vertex(55,68){2}

\Curve{(203,72)(213,75)(223,85)}
\Curve{(260,100)(265,92)(276,88)}
\Curve{(276,88)(277,84)(283,73)(288,72)}
\Vertex(223,85){1.3}
\Vertex(260,100){1.7}
\Vertex(276,88){1.3}
\Vertex(288,72){2}
\Line(203,72)(195,75)
\Line(203,72)(195,67)
\Vertex(203,72){1}
\Vertex(195,75){2}
\Vertex(195,67){2}

\Curve{(205,68)(213,65)(220,54)}
\Curve{(220,54)(232,49)(240,40)}
\Curve{(271,54)(280,64)(292,68)}
\Vertex(220,54){1.3}
\Vertex(240,40){1.7}
\Vertex(271,54){1.3}
\Line(292,68)(300,73)
\Line(292,68)(300,65)
\Vertex(292,68){1}
\Vertex(300,73){2}
\Vertex(300,65){2}
\Vertex(205,68){2}


\Vertex(211,71){1.3}
\Vertex(247,90){1.3}
\Vertex(280,72){1.3}

\Vertex(213,68){1.3}
\Vertex(253,52){1.3}
\Vertex(284,69){1.3}

\Text(97,25)[]{(a)}
\Text(247,25)[]{(b)}

\end{picture}
}}
\hspace{-5mm}
\scalebox{0.78}{\mbox{
\begin{picture}(151,115)(44,20)

\Curve{(53,72)(63,75)(73,85)}
\Curve{(73,85)(80,85)(90,93)}
\Curve{(90,93)(97,93)(110,100)}
\Curve{(110,100)(115,92)(126,88)}
\Curve{(126,88)(127,84)(133,73)(138,72)}
\Vertex(73,85){1.3}
\Vertex(90,93){1.3}
\Vertex(110,100){1.7}
\Vertex(126,88){1.3}
\Vertex(138,72){2}
\Line(53,72)(45,75)
\Line(53,72)(45,67)
\Vertex(53,72){1}
\Vertex(45,75){2}
\Vertex(45,67){2}

\Curve{(55,68)(63,65)(70,54)}
\Curve{(70,54)(82,49)(90,40)}
\Curve{(90,40)(95,45)(106,47)}
\Curve{(106,47)(112,53)(121,54)}
\Curve{(121,54)(130,64)(142,68)}
\Vertex(70,54){1.3}
\Vertex(90,40){1.7}
\Vertex(106,47){1.3}
\Vertex(121,54){1.3}
\Line(142,68)(150,73)
\Line(142,68)(150,65)
\Vertex(142,68){1}
\Vertex(150,73){2}
\Vertex(150,65){2}
\Vertex(55,68){2}

\Text(97,25)[]{(c)}

\end{picture}
}}

\caption{(a) In a hard gluon-gluon subcollision the outgoing gluons
will be colour-connected to the projectile and target remnants.
Initial state radiation may give extra gluon kinks, which are ordered 
in rapidity. (b) A second hard scattering would naively be expected to give
two new strings connected to the remnants. (c) In the fits to data the gluons
are colour reconnected, so that the total string length becomes as short as 
possible.}
\label{fig:recoupl}
\end{figure}

Although the \textsc{Pythia} model has been tuned to fit most of the 
CDF data on minimum bias and underlying events, 
there are a number of open questions:

- How does a many-parton system hadronize in an event with multiple hard
subcollisions? The result expected in the string hadronization model
is illustrated in fig.~\ref{fig:recoupl}.
In a hard gluon-gluon subcollision the outgoing gluons
will be colour-connected to the projectile and target remnants, as shown in
fig.~\ref{fig:recoupl}a.
Initial state radiation may give extra gluon ``kinks'', which are ordered 
in rapidity. A second hard scattering would naively be expected to give
two new strings connected to the remnants as in fig.~\ref{fig:recoupl}b. 
As a result this would give almost doubled multiplicity. This is not in 
agreement with data. In the successful fits \cite{Field} it is instead assumed 
that the gluons are colour reconnected, so that the total string length becomes 
as short as possible,
see fig.~\ref{fig:recoupl}c. This colour reconnection implies that a minimum 
number of hadrons share the transverse momentum of the partons.



- In most analysis, including \textsc{Pythia}, it is assumed that the impact 
parameter and momentum distributions of the interacting partons factorize. 
In a dipole cascade model strong non-factorizing correlations are expected 
between impact parameter, momentum, and multiplicity of the colliding
partons in a proton. Besides such correlations, also the expected
large fluctuations in multiplicity and impact parameter distributions 
are more difficult to include in a momentum space formalism.

\section{Eikonal formalism}

A formalism in transverse coordinate space is very suitable for describing
rescattering 
and multiple collisions. In a process where a particle undergoes 
successive interactions with transverse momenta $\mathbf{k}_{\perp i}$,
the resulting transverse momentum $\mathbf{k}_\perp=\sum \mathbf{k}_{\perp i}$ 
is given by a convolution of the different interactions. As the Fourier 
transform of a convolution is given by a simple product, we see that
in impact parameter space the multiple interactions are described
by a product of the $S$-matrix elements for the individual interactions:
\begin{equation}
S(b)=S_1(b) S_2(b) S_3(b).
\end{equation}
Thus for $S_i=e^{-\eta_i(b)}$ we find $S=e^{-\sum \eta_i}$.

\subsection{Weizs\"acker-Williams method of virtual quanta}

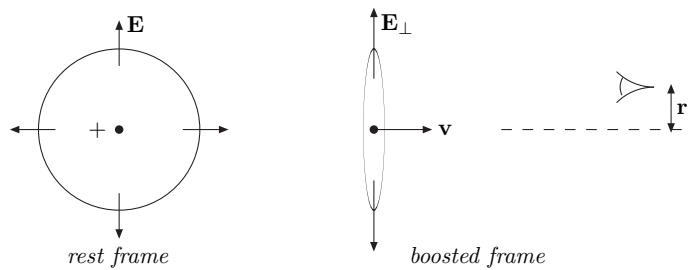
\begin{figure}
\begin{center}
\scalebox{0.80}{\mbox{
\begin{picture}(360,140)(0,0)
\BCirc(70,70){38}
\Vertex(70,70){2}
\LongArrow(40,70)(20,70)
\LongArrow(100,70)(120,70)
\LongArrow(70,100)(70,120)
\LongArrow(70,40)(70,20)
\Text(74,120)[l]{$\mathbf{E}$}
\Text(60,70)[]{+}

\Oval(190,70)(38,5)(0)
\Vertex(190,70){2}
\LongArrow(190,94)(190,126)
\LongArrow(190,46)(190,14)
\LongArrow(190,70)(216,70)
\Text(194,120)[l]{$\mathbf{E}_\perp$}
\Text(221,70)[l]{$\mathbf{v}$}

\DashLine(250,70)(335,70){5}
\CArc(322,65)(25,90,135)
\CArc(322,115)(25,225,270)
\CArc(315,90)(9,150,205)

\LongArrow(330,80)(330,90)
\LongArrow(330,80)(330,70)
\Text(334,80)[l]{$\mathbf{r}$}

\Text(70,10)[]{\emph{rest frame}}
\Text(240,10)[]{\emph{boosted frame}}

\end{picture}
}}
\end{center}
\caption{A Coulomb field in a boosted frame is compressed to a flat pancake}
\label{fig:WW}
\end{figure}

A Coulomb field which is boosted is contracted to a flat pancake, with a 
dominantly transverse electric field
\begin{equation}
\mathbf{E}_\perp \sim g\frac{\mathbf{r}}{r^2},
\label{eq:Efield}
\end{equation}
and an orthogonal transverse
magnetic field with the same magnitude. Here $\mathbf{r}$ is the 
(two-dimensional) distance between the position of the central charge and
the point of observation (see fig. \ref{fig:WW}).
The pulse will be very short in time, and can be approximated by a 
$\delta$-function:
\begin{equation}
I(t) \sim E_\perp B_\perp \sim g^2 \frac{1}{r^2} \delta(t).
\end{equation}
The frequency distribution is given by the Fourier transform of $I(t)$,
which is a constant. Thus the distribution of photons, or gluons, seen by 
an observer at point $\mathbf{r}$ is given by
\begin{equation}
dn \sim g^2\frac{d^2r}{r^2}\frac{d\omega}{\omega} \sim 
g^2\frac{d^2q_\perp}{q_\perp^2}\frac{d\omega}{\omega}.
\end{equation}


Inside a proton there is a very complicated colour field. 
We may however imagine that within some distance
$r_0$ from a colour charge it is approximately a Coulomb field, while for
larger distances the charge is screened by other charges. If $r_0$ is around
0.1 fm this would give an effective cutoff for hard subcollisions with
$p_{\perp \mathrm{min}} \approx 10\, \mathrm{fm}^{-1} \approx 2$ GeV, as 
obtained in the fits to experimental data.

A bremsstrahlung gluon from this field will change the colour of the
initial charge. If \eg the initial charge is a red quark, it may emit
a red-antiblue gluon and change its colour to blue. The result is that the
initial red Coulomb field now terminates at the gluon charge, and
a blue-antiblue dipole field is formed between the quark and the gluon.
The emission of softer gluons now get two separate contributions, one 
from the modified Coulomb field and one from the new colour dipole between the 
quark and the gluon. The repeated emission of more gluons results in a 
cascade, as discussed in the next section.


\subsection{The Mueller dipole model}

\begin{figure}
\scalebox{0.8}{\mbox{
\begin{picture}(125,72.5)(-10,-20)
\Line(5,5)(100,5)
\Line(5,5)(38,45.5)
\Line(38,45.5)(100,5)
\Vertex(5,5){2}
\Vertex(100,5){2}
\Vertex(38,45.5){2}
\LongArrow(38,45.5)(60,67.5)
\LongArrow(38,45.5)(56.3,34)
\LongArrow(38,45.5)(78.3,56)
\Text(80,59)[lt]{$\mathbf{E}_\perp$}
\Text(35.5,45)[rb]{$\mathbf{r}$}
\Text(50,2.5)[t]{$R$}
\Text(21.5,25.5)[rb]{$r_1$}
\Text(76.5,25)[b]{$r_2$}
\Text(1,5)[r]{$red$}
\Text(104,5)[l]{$\overline{red}$}
\Text(65,-18)[]{{\large (a)}}
\end{picture}
}}
\hspace{8mm}
\scalebox{0.65}{\mbox{
\begin{picture}(340,80)(0,-30)
\Vertex(10,80){2}
\Vertex(10,0){2}
\Text(5,80)[]{$\mathrm{x}$}
\Text(5,0)[]{$\mathrm{y}$}
\Line(10,80)(10,0)
\LongArrow(30,40)(60,40)
\Vertex(100,80){2}
\Vertex(100,0){2}
\Vertex(120,50){2}
\Text(95,80)[]{$\mathrm{x}$}
\Text(95,0)[]{$\mathrm{y}$}
\Text(128,50)[]{$\mathrm{z}$}
\Line(100,80)(120,50)
\Line(120,50)(100,0)
\LongArrow(140,40)(170,40)
\Vertex(205,80){2}
\Vertex(225,50){2}
\Vertex(233,30){2}
\Vertex(205,0){2}
\Text(200,80)[]{$\mathrm{x}$}
\Text(200,0)[]{$\mathrm{y}$}
\Text(233,50)[]{$\mathrm{z}$}
\Text(241,30)[]{$\mathrm{w}$}
\Line(205,80)(225,50)
\Line(225,50)(233,30)
\Line(233,30)(205,0)
\LongArrow(255,40)(285,40)
\Line(310,80)(320,70)
\Line(320,70)(302,63)
\Line(302,63)(325,58)
\Line(325,58)(330,50)
\Line(330,50)(320,43)
\Line(320,43)(338,30)
\Line(338,30)(310,35)
\Line(310,35)(310,0)
\Vertex(310,80){2}
\Vertex(320,70){2}
\Vertex(302,63){2}
\Vertex(325,58){2}
\Vertex(330,50){2}
\Vertex(320,43){2}
\Vertex(338,30){2}
\Vertex(310,35){2}
\Vertex(310,0){2}
\Text(170,-29)[]{{\large (b)}}
\end{picture}
}}
\caption{(a) The transverse colour-electric field in a colour dipole.
(b) Gluon emission splits the dipole into two dipoles. Repeated
emissions give a cascade, which produces a chain of dipoles.}
\label{fig:dipolesplit}
\end{figure}
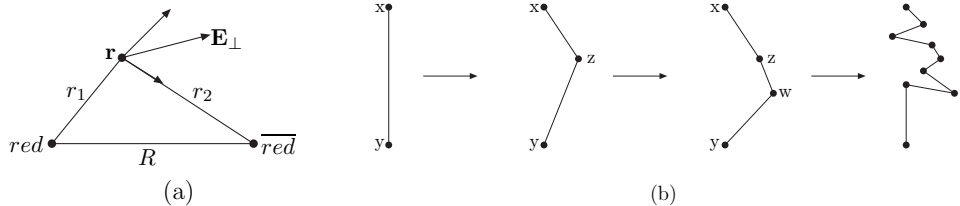

A dipole cascade model in transverse coordinate space was developed
by A. Mueller in a series of papers \cite{Mueller}.
We study an initial colour neutral quark-antiquark system.
The colour dipole field gets two contributions of the form 
in eq.~(\ref{eq:Efield}), as shown in fig.~\ref{fig:dipolesplit}a. Adding 
these with opposite signs the resulting 
transverse colour-electric field $\mathbf{E}_\perp$
is given by
\begin{equation}
E_\perp^2\sim g^2\frac{R^2}{r_1^2\! r_2^2},
\label{eq:split}
\end{equation}
where $R$ is the size of the parent dipole, and $r_1$ and $r_2$ are the 
distances from the point $\mathbf{r}$ to the two charges (see 
fig.~\ref{fig:dipolesplit}a).
As discussed above this implies that the probability to emit a gluon
in the point $\mathbf{r}$ becomes
\begin{equation}
\frac{d\mathcal{P}}{dy}=\bar{\alpha}\frac{d^2 r}{2\pi}
     \frac{R^2}{r_1^2 \,r_2^2}.
\label{eq:dipolesplit}
\end{equation}
If the initial charges were \eg red and antired, the emitted gluon may be 
red-antiblue, changing the initially red charge to blue. Therefore such an 
emission implies that the dipole is split in two connected dipoles, one 
blue-antiblue and one red-antired. These can then also split repeatedly 
in a dipole cascade, as shown in fig.~\ref{fig:dipolesplit}b.
In ref.~\cite{Mueller} it is demonstrated that this cascade reproduces the
LL BFKL evolution.



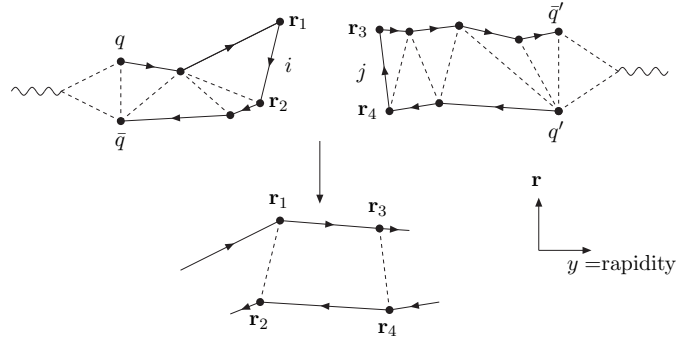
\begin{figure}
\begin{center}
\scalebox{0.75}{\mbox{
\begin{picture}(400,175)(0,7)
\Photon(35,130)(60,130){2}{3}
\Photon(365,140)(340,140){2}{3}
\DashLine(60,130)(90,145){2}
\DashLine(60,130)(90,115){2}
\DashLine(340,140)(310,160){2}
\DashLine(340,140)(310,120){2}
\DashLine(90,145)(90,115){2}
\DashLine(310,120)(310,160){2}
\DashLine(90,115)(120,140){2}
\DashLine(120,140)(145,118){2}
\DashLine(160,124)(120,140){2}
\DashLine(310,120)(290,156){2}
\DashLine(310,120)(260,163){2}
\DashLine(260,163)(250,124){2}
\DashLine(250,124)(235,160){2}
\DashLine(235,160)(225,120){2}
\ArrowLine(90,145)(120,140)
\ArrowLine(120,140)(170,165)
\ArrowLine(145,118)(90,115)
\ArrowLine(170,165)(160,124)
\ArrowLine(160,124)(145,118)
\ArrowLine(120,140)(170,165)
\ArrowLine(290,156)(310,160)
\ArrowLine(260,163)(290,156)
\ArrowLine(310,120)(250,124)
\ArrowLine(225,120)(220,161)
\ArrowLine(250,124)(225,120)
\ArrowLine(220,161)(235,160)
\ArrowLine(235,160)(260,163)
\Vertex(310,160){2}
\Vertex(310,120){2}
\Vertex(90,145){2}
\Vertex(90,115){2}
\Vertex(120,140){2}
\Vertex(170,165){2}
\Vertex(145,118){2}
\Vertex(160,124){2}
\Vertex(290,156){2}
\Vertex(260,163){2}
\Vertex(310,120){2}
\Vertex(250,124){2}
\Vertex(235,160){2}
\Vertex(260,163){2}
\Vertex(220,161){2}
\Vertex(225,120){2}
\Text(90,155)[]{$q$}
\Text(90,105)[]{$\bar{q}$}
\Text(310,170)[]{$\bar{q}'$}
\Text(310,110)[]{$q'$}
\Text(210,161)[]{$\mathbf{r}_3$}
\Text(215,120)[]{$\mathbf{r}_4$}
\Text(180,165)[]{$\mathbf{r}_1$}
\Text(170,124)[]{$\mathbf{r}_2$}
\Text(212,140)[]{$j$}
\Text(175,144)[]{$i$}
\LongArrow(190,105)(190,75)
\DashLine(170,65)(160,24){2}
\ArrowLine(160,24)(145,18)
\ArrowLine(120,40)(170,65)
\DashLine(225,20)(220,61){2}
\ArrowLine(250,24)(225,20)
\ArrowLine(220,61)(235,60)
\Vertex(220,61){2}
\Vertex(225,20){2}
\Vertex(170,65){2}
\Vertex(160,24){2}
\ArrowLine(170,65)(220,61)
\ArrowLine(225,20)(160,24)
\Text(170,75)[]{$\mathbf{r}_1$}
\Text(160,14)[]{$\mathbf{r}_2$}
\Text(220,71)[]{$\mathbf{r}_3$}
\Text(225,10)[]{$\mathbf{r}_4$}

\LongArrow(300,50)(300,75)
\LongArrow(300,50)(325,50)
\Text(300,85)[]{$\mathbf{r}$}
\Text(343,43)[]{$y=$rapidity}
\end{picture}
}}
\end{center}
\caption{\label{fig:dipcoll} A symbolic picture of a $\gamma^* \gamma^*$
    collision in $y-\mathbf{r}_\perp$-space. When two colliding dipoles 
    interact via gluon exchange the colors are reconnected forming dipole 
    chains stretched between the remnants of the colliding systems.}
\end{figure}

When two such cascades collide as in fig.~\ref{fig:dipcoll}, two dipoles ($i$ 
and $j$ with endpoints ($\mathbf{r}_1,\mathbf{r}_2$) and
($\mathbf{r}_3,\mathbf{r}_4$) respectively) may interact via gluon exchange.
Adding coherently the exchange between charge or anticharge in
both dipoles gives the interaction probability 
\begin{equation}
f_{ij}= \frac{\alpha_s^2}{8} 
\ln^2\left( \frac{(\mathbf{r}_1-\mathbf{r}_3)^2 (\mathbf{r}_2-\mathbf{r}_4)^2}
{(\mathbf{r}_1-\mathbf{r}_4)^2 (\mathbf{r}_2-\mathbf{r}_3)^2} \right).
\label{eq:intprob}
\end{equation}
As the exchanged gluon carries colour, the dipole chains 
are reconnected, as also indicated in fig.~\ref{fig:dipcoll}. The result is two
dipole chains connecting the remnants of the projectile and the target systems,
as also illustrated in fig.~\ref{fig:recoupl}.

It is also possible that several pairs ($i,j$) interact.
In the eikonal approximation the total, diffractive, and elastic cross sections
are given by

\begin{equation}
\left\{ \begin{array}{l}
\sigma_{\mathrm{tot}} \sim d^2b\, 2\langle ( 1-e^{-\sum f_{ij}})\rangle, \\
\sigma_{\mathrm{diff}}\! \sim d^2b\, \langle(1-e^{-\sum f_{ij}})^2\rangle \,\,\,\,\,
 \mathrm{(incl.\,\, elastic)},\\
\sigma_{\mathrm{el}}\,\, \sim d^2b\, (\langle 1-e^{-\sum f_{ij}}\rangle)^2.
\label{eq:eikonalcrossections}
\end{array} \right.
\end{equation}
We note that as the parentheses always are smaller than unity, unitarity
is always satisfied. We also note that diffractive excitation, which is given
by the difference between the second and third lines in 
eq.~(\ref{eq:eikonalcrossections}), is directly determined by the fluctuations
in the cascade evolution.

\begin{figure}
\begin{center}
\includegraphics[scale=0.45]{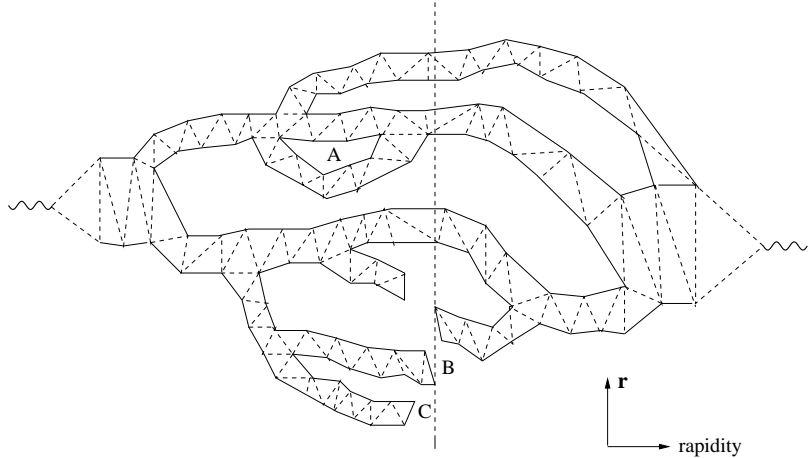}
\end{center}
\caption{A schematic picture of a collision between two evolved dipole 
cascades. In the cms (indicated by a dashed line) there are three separate 
subcollisions.
There is also a dipole loop (marked $A$) within the left system, caused by a
``dipole swing''. The parts of the evolutions marked $B$ and $C$ do not 
interact, and should be treated as virtual.}
\label{fig:finalstate}
\end{figure}

A schematic picture of a collision between two evolved dipole cascades is 
shown in fig.~\ref{fig:finalstate}. In this example there are in the cms 
(indicated by a dashed line) three separate subcollisions. These also 
correspond to the exchange of three pomerons. They result
in two closed loops formed by chains of colour dipoles, in addition to the 
two dipole chains which connect the projectile and target remnants. 
Fig.~\ref{fig:finalstate} also contains a dipole loop (denoted $A$) inside 
the evolution of the left system, before it collides. Such loops are, however,
not included in Mueller's model, which only accounts for loops cut
in the specific Lorentz frame used for the calculation. This implies that 
this model is not Lorentz frame independent.

\subsection{Problems}

Although Mueller's cascade model, and also \eg the Balitsky-Kovchegov evolution 
equation, reproduce LL BFKL evolution and satisfies 
the unitarity constraints, they have a set of problems:

\newpage
\begin{itemize}

\item LL BFKL is not good enough. NLL corrections are very large.

\item Non-linear effects in the evolution are not included.

\item Massless gluon exchange implies a violation of Froissart's bound.

\item It is difficult to include fluctuations and correlations; 
the BK equation represents a mean field approximation.

\item They can only describe inclusive features, and not the production of
exclusive final states.

\item Analytic calculations are mainly applicable at extreme energies,
well beyond what can be reached experimentally.

\end{itemize}

\subsection{Monte Carlo simulations}

Non-leading effects, \eg effects of energy conservation and a running
coupling, are often easier included in Monte Carlo simulations. 
A simulation of Muller's cascade model has been presented by 
Salam \cite{SalamMC}.
However, as the dipole splitting diverges for small dipole sizes, $r$, 
(see eq.~(\ref{eq:dipolesplit})) this
simulation is hampered by numerical difficulties. It is noticed that 
the fluctuations in the number of dipoles is very large, which causes 
serious problems.

\section{A new approach}

In refs.~\cite{newmodel, elastic} we describe a modification of Mueller's 
cascade model with the following features:
\begin{itemize}
\vspace{-2mm}
\item It includes essential NLL BFKL effects.
\vspace{-2mm}
\item It includes non-linear effects in the evolution.
\vspace{-2mm}
\item It includes essential correlations and fluctuations.
\vspace{-2mm}
\item It can also describe exclusive final states.
\vspace{-2mm}
\end{itemize}

The model is also implemented in a Monte Carlo simulation program.
Here the NLL effects significantly reduce the production of small dipoles,
and thereby also the associated numerical difficulties with very large 
dipole multiplicities are avoided. Confinement corrections are also included, 
but this is numerically less important than NLL and non-linear effects.

\subsection{NLL effects}

The NLL corrections to BFKL evolution have three major sources
\cite{Salamnll}:

\begin{itemize}
\vspace{-2mm}

\item \emph{The running coupling.}
\vspace{-1mm}

This is relatively easily included in a MC simulation process.
\vspace{-2mm}
\item \emph{Non-singular terms in the splitting function.}
\vspace{-1mm}

These terms suppress large $z$-values in the individual branchings, and 
prevent the daughter from being faster than her recoiling parent. 
Most of this effect is taken care of by including energy-momentum conservation
in the evolution.

\vspace{-2mm}
\item \emph{Projectile-target symmetry.}
\vspace{-1mm}

This is also called energy scale terms, and is essentially equivalent
to the so called consistency constraint. This effect is taken into account by 
conservation of both positive and negative lightcone momentum components,
$p_+$ and $p_-$.
\vspace {-2mm}
\end{itemize}

The treatment of these effects includes effects beyond NLL, in a way similar to
the treatment by Salam in ref.~\cite{Salamnll}. Thus the power 
$\lambda_{\mathrm{eff}}$,
determining the growth for small $x$, is not negative for large
values of $\alpha_s$.

\subsection{Non-linear effects and saturation}

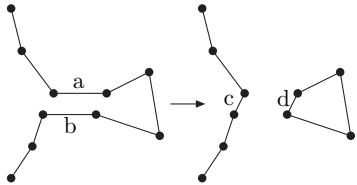
\begin{figure}
\begin{center}
\scalebox{0.8}{\mbox{
\begin{picture}(180,91)(0,9)
\Line(10,10)(20,25)
\Line(25,40)(20,25)
\Line(25,40)(50,40)
\Line(50,40)(80,30)
\Line(75,60)(80,30)
\Line(75,60)(55,50)
\Line(30,50)(55,50)
\Line(30,50)(15,70)
\Line(15,70)(10,90)
\LongArrow(85,45)(100,45)
\Line(100,10)(110,25)
\Line(115,40)(110,25)
\Line(115,40)(120,50)
\Line(140,40)(170,30)
\Line(165,60)(170,30)
\Line(165,60)(145,50)
\Line(145,50)(140,40)
\Line(120,50)(105,70)
\Line(105,70)(100,90)
\Vertex(10,10){2}
\Vertex(20,25){2}
\Vertex(25,40){2} 
\Vertex(50,40){2} 
\Vertex(80,30){2} 
\Vertex(75,60){2}
\Vertex(55,50){2} \Vertex(30,50){2} \Vertex(15,70){2} \Vertex(10,90){2}
\Vertex(100,10){2} \Vertex(110,25){2} \Vertex(115,40){2} \Vertex(120,50){2}
\Vertex(140,40){2} \Vertex(170,30){2} \Vertex(165,60){2} \Vertex(145,50){2}
\Vertex(105,70){2} \Vertex(100,90){2}
\Text(38,35)[]{b} \Text(42,55)[]{a} \Text(113,47)[]{c} \Text(139,48)[]{d} 
\end{picture}
}}
\end{center}
\caption{\label{fig:swing} Schematic picture of a dipole swing. 
  If the two dipoles $a$ and $b$ have the same color, they can be replaced
  by the dipoles $c$ and $d$. The result is a closed loop formed within an 
  individual dipole cascade.}  
\end{figure}

As mentioned above, dipole loops within the evolution, as indicated by
the letter $A$ in fig.~\ref{fig:finalstate}, are not included in Mueller's
cascade model or in the JIMWLK or BK equations. Like for dipole scattering
the probability for such loops is given by $\alpha_s$, and therefore
formally colour suppressed compared to dipole splitting, which is
proportional to $\bar{\alpha}=N_c \alpha_s/\pi$. These loops are therefore
related to the probability that two dipoles have the same colour. Two dipoles
with the same colour form a quadrupole. Such a field may be better
approximated by two dipoles formed by the closest colour-anticolour
charges. This corresponds to a recoupling of the colour dipole chains,
as indicated in fig.~\ref{fig:swing}. We call this process a dipole ``swing''.
Although this mechanism does not give an explicitly frame independent result, 
MC simulations show that it is a very good approximation 
(see also fig.~\ref{fig:sigmatot} below).
We note that in this formulation the number of dipoles is not reduced.
The saturation effect is obtained because the swing favours the formation
of smaller dipoles, which have a smaller cross section. Thus in an evolution
in momentum space the swing would not correspond to an absorption of gluons 
below the saturation line $k_\perp^2 = Q_s^2(x)$; it would rather correspond 
to lifting the gluons to higher $k_\perp$ above this line.


\subsection{Confinement}

The exchange of massless gluons gives an interaction of infinite range, which
eventually will violate Froissart's bound. This long range interaction is 
suppressed by
introducing an effective gluon mass. Thus the gluon propagater $1/k_\perp^2$
is replaced by $1/(k_\perp^2 + m^2)$. This implies that 
the expression for the transverse electric field in eq.~(\ref{eq:Efield})
should be replaced by $\mathbf{r}/(r r_{\mathrm{max}}) K_1(r/ r_{\mathrm{max}})$,
and in the interaction probability in eq.~(\ref{eq:intprob}) the logarithms
$\ln(1/r)$ are replaced by $K_0(r_{\mathrm{max}}/r)$.
Here $r_{\mathrm{max}}=1/m$ and
$K_1$ and $K_0$ are modified Bessel functions.
For small values of $r$ the expressions are unchanged, but when $r$ becomes
larger than $r_{\mathrm{max}}$ they become exponentially suppressed.

\subsection{Exclusive final states}

As the size of a dipole is also associated with (the inverse of) its 
transverse momentum, the event depicted in fig.~\ref{fig:finalstate} can also
be interpreted as an exclusive final state with definite parton momenta. 
It is true that taking the Fourier transform is not exactly the same as 
replacing $\mathbf{r}$ by $\mathbf{r}/r^2$. Therefore, when there is a
conflict, what we really are generating is perhaps not $r$ with the
approximation $k_\perp\approx 1/r$, but rather generating $k_\perp$ with
$r\approx1/k_\perp$. There 
are also some other problems which have to be solved. In the evolution of 
a particle with a high 
value for the lightcone momentum, $p_+$, but very little $p_-$,
the gluons in the cascade will accumulate a large deficit of $p_-$,
which has to be ``payed back'' when it collides with a target having a 
large $p_-$ but small $p_+$. Dipole chains like the ones indicated by 
$B$ and $C$ in fig.~\ref{fig:finalstate}, which do not interact,
can not come on shell and 
have to be treated as virtual. Thus only gluons which are connected to 
daughters involved in the collisions should  be included in the initial 
state radiation.

An important feature is that all dipoles are connected in chains, which 
implies strong non-trivial correlations \cite{emilcorr}. A toy model 
calculation by R. Corke \cite{Corke} shows that such correlations can have a 
big effect on the need for colour reconnections in the final state.

The full calculation of the final state properties is not yet implemented
in the MC, but will be ready in the near future.

\section{Applications}

For the results presented in this section we have used a simple model for 
the proton wave function, in form of three dipoles in an 
equilateral triangle. For larger $Q^2$ the wavefunction for a virtual 
photon is determined 
by perturbation theory, but for smaller $Q^2$ it is necessary to include a
nonperturbative component representing the soft hadronic part of the photon. 
For more details see ref.~\cite{elastic}.

\subsection{Total cross sections}

\begin{figure}
  \includegraphics[angle=270, scale=0.50]{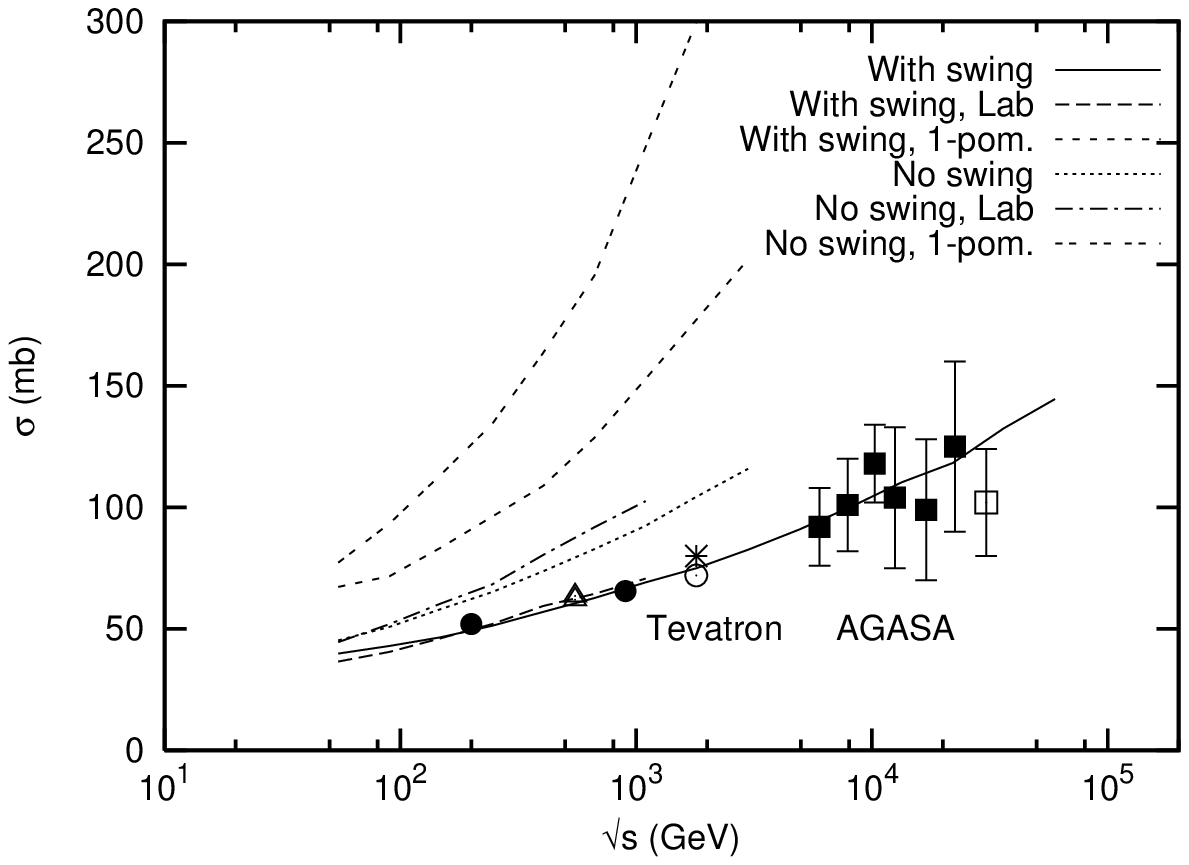}
  \includegraphics[angle=270,  scale=0.42]{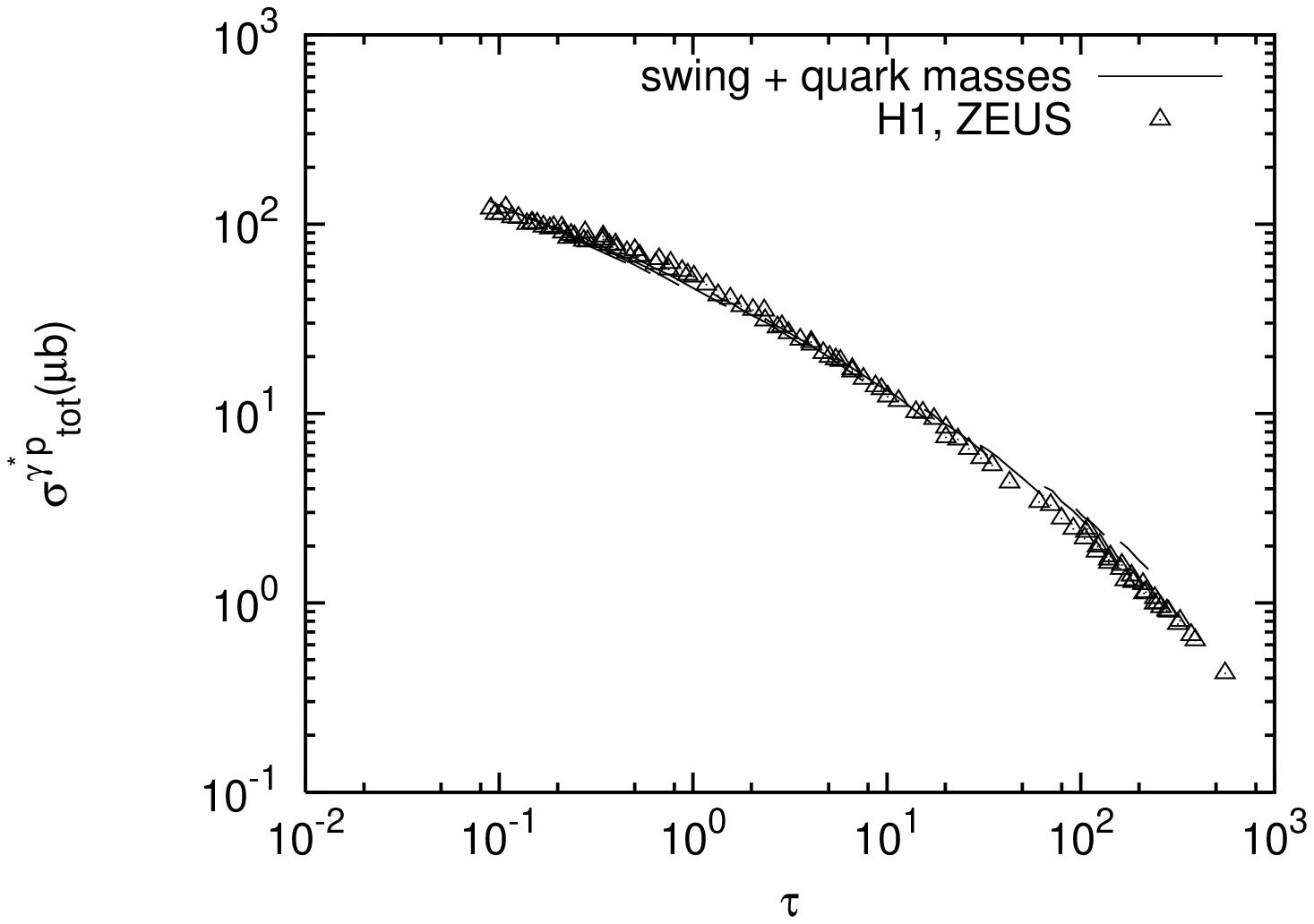}
  \caption{\emph{Left}: The model results for the total $pp$ scattering 
   cross section. Results are presented
    for evolution with and without the dipole swing mechanism. The
    one pomeron result and the result obtained in a frame where one of 
    the protons is almost at rest is also shown.
    \emph{Right}:  $\gamma^* p$ total cross section compared to data from 
    H1 \cite{Adloff:2000qp} and 
    Zeus \cite{Breitweg:2000yn}. The result is plotted as a
     function of the GBW scaling variable $\tau=Q^2/Q_{s}^2(x)$.
  }
   \label{fig:sigmatot}
\end{figure}

The total $pp$ cross section is shown in fig.~\ref{fig:sigmatot}a. 
We note that the one-pomeron result, which neglects unitarity
constraints, is about a factor four too high at Tevatron energies. Also
saturation effects inside the evolution, simulated by the swing, 
have a significant effect, reducing the cross section by about 30\%. 
We also see that including the swing makes the result obtained in 
the rest frame of the target (long dashed line) very close to the result
in the cms (solid line), which shows that the result is almost Lorentz
frame independent.

The total $\gamma^* p$ cross section is presented in fig.~\ref{fig:sigmatot}b
against the Golec-Biernat--W\"{u}sthoff scaling parameter $\tau=Q^2/Q_s^2$,
where $Q_{s}^2=Q_0^2 (x_0/x)^\lambda$ with $Q_0=1\mathrm{GeV}$, 
$x_0=3\cdot 10^{-4}$, $\lambda=0.29$ \cite{GBW}. 
We note that our result, in agreement with the data,
satisfies geometric scaling both for small $\tau$, where the result is
influenced by saturation, and for larger $\tau$ where saturation 
is not essential.

\subsection{Elastic and quasielastic cross sections}

\begin{figure}
\begin{center}
\includegraphics[scale=1.3]{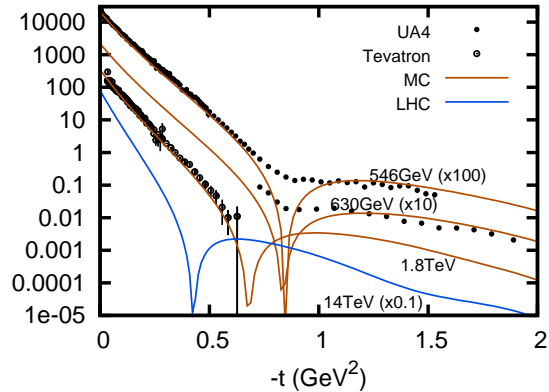}
\end{center}
\caption{The differential cross section for $pp$ elastic scattering, 
together with data from UA4 \cite{UA4} and CDF \cite{CDFelastic}.
The figure also includes a prediction for the LHC.}
\label{fig:pptdep}
\end{figure}

Fig.~\ref{fig:pptdep} shows the elastic $pp$ cross section. As the real part 
of the amplitude is neglected the diffractive dip is a real zero. The position 
of this dip can be tuned at one energy, but the change to smaller $t$
at higher energies is a result of the evolution, which cannot be tuned.

\begin{figure}
  \includegraphics[scale=1]{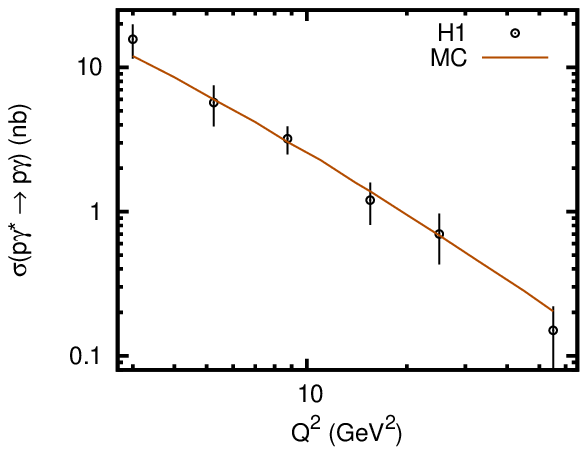}
  \includegraphics[scale=1]{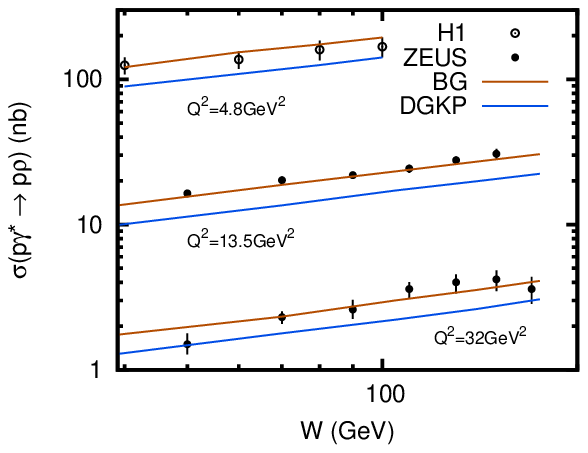}
  \caption{\emph{Left}: The cross section for $\gamma^* p \rightarrow \gamma p$
  for $W$ = 82 GeV as function of $Q^2$, compared to data from H1 
  \cite{dvcs}.
    \emph{Right}: The cross section for $\gamma^* p \rightarrow \rho p$
  for $Q^2$~= 4.8, 13.5, and 32 $\mathrm{GeV}^2$ as function of $W$, compared 
  to data from H1 \cite{H1rho} and ZEUS \cite{ZEUSrho}. Two different 
  wavefunctions have been used for the $\rho$-meson, boosted Gaussian
  (solid line) and DGKP (dashed line).}
   \label{fig:quasi}
\end{figure}

Quasielastic $\gamma^* p$ scattering is also well reproduced by the model,
including the dependence on $Q^2$, $W$, and $t$.
As two examples fig.~\ref{fig:quasi} shows
the $Q^2$-dependence for DVCS and the $W$-dependence for 
$\gamma^* p \rightarrow \rho p$. For the latter case two different
$\rho$-meson wavefunctions were used, a boosted Gaussian wavefunction
\cite{boostedgaussian} and the DGKP \cite{DGKP} wavefunction. Both models
show similar growth with $W$, but our normalization
agrees better with the boosted Gaussian wavefunction.

\subsection{Diffraction}

\begin{figure}
  \includegraphics[scale=1]{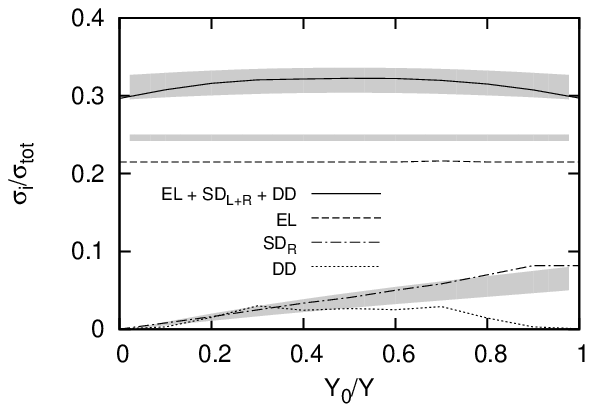}
  \includegraphics[scale=1]{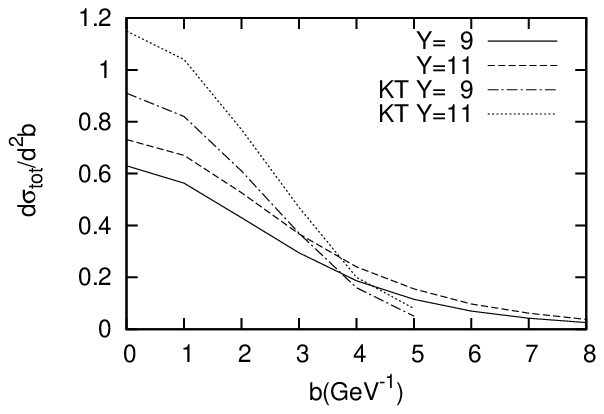}
  \caption{\emph{Left}: The ratio between the total diffractive and the total
cross sections (solid line) together with the contribution from elastic 
(dashed), single-right (dash-dotted), and double diffractive (dotted) 
cross sections at 1.8 TeV. The shaded bands are estimates from CDF data
\cite{CDFdiff}. \emph{Right}: The impact parameter profile for dipole-proton 
collisions for a initial dipole size $r=2\,\mathrm{GeV}^{-1} \approx 0.4\,$fm
at two different energies, $Y=\ln s=$ 9 and 11. Our results (solid and dashed 
lines) are compared to those from the Kowalski-Teaney \cite{KT} model
(dot-dashed and dotted lines)}
   \label{fig:diff}
\end{figure}

As discussed above diffractive excitation is directly determined by the
fluctuations in the evolution. As an example, if we calculate the expression
\begin{equation}
\int d^2 b \{\langle \langle 1-e^{-\sum f_{ij}} \rangle_L^2 \rangle_R -
\langle 1-e^{-\sum f_{ij}} \rangle_{LR}^2 \}
\end{equation}
in a frame where the right-moving proton is evolved $y_R$ rapidity units,
it gives the cross section for single diffractive excitation of this proton
to masses satisfying $M_X^2 < M_0^2 \exp(y_R)$, with $M_0 \sim 1$ GeV.
Here $L$ and $R$ indicate averages over left- and right-moving proton
cascades respectively.
The result obtained when varying $y_R$ is shown in fig.~\ref{fig:diff}a
compared with corresponding results from the Tevatron.

We also note that as the 
fluctuations in the evolution are quite large, less fluctuations are 
needed in the impact parameter profile. Thus this profile is more gray, 
and less black and white, than in fits where the fluctuations in the 
evolution are not taken into account. This is seen in fig.~\ref{fig:diff}b,
which shows a comparison between our model
and the result by Kowalski and Teaney \cite{KT} for the scattering of a 
dipole against a proton. In the latter analysis 
the fluctuations in the evolution are not included.


\section{Summary}

\begin{itemize}

\item A new dipole formulation of high energy collisions in transverse 
coordinate space is presented

\item It has the following main ingredients:

  \begin{itemize}
\vspace{-1mm}

  \item NLL corrections to BFKL

  \item Non-linear effects: saturation and multiple subcollisions

  \item Confinement effects
 
  \item Includes momentum-impact parameter correlations

  \item Simple proton and photon models

  \item MC implementation
\vspace{-1mm}

  \end{itemize}
\vspace{-1mm}

\item It gives a fair description of data for:

  \begin{itemize}
\vspace{-1mm}

  \item Total cross sections for $pp$ and $\gamma^* p$ collisions

  \item (Quasi-)elastic scattering in $pp$ and $\gamma^* p$

  \item Diffractive excitation
\vspace{-1mm}
  \end{itemize}

\item To come soon: Generation of exclusive final states

\item Wanted: Better understanding of the connection to the $t$-channel picture
of pomeron loops

\end{itemize}


\end{document}